\documentclass{PoS}

\usepackage{amsmath,amssymb}
\usepackage{wrapfig}

\title{Crystalline Confinement}

\ShortTitle{Crystalline Confinement}

\author{\speaker{D. Banerjee, P. Widmer}\\
       Albert Einstein Center for Fundamental Physics,\\
       Institute for Theoretical Physics,\\ 
       Bern University, Switzerland\\
        E-mail: \email{dbanerjee@itp.unibe.ch,widmer@itp.unibe.ch}}

\author{F.-J. Jiang\\
        Department of Physics, National Taiwan Normal University 88,\\
        Sec. 4, Ting-Chou Rd., Taipei 116, Taiwan\\
        E-mail: \email{fjjiang@ntnu.edu.tw}}

\author{U.-J. Wiese\\
       Albert Einstein Center for Fundamental Physics,\\
       Institute for Theoretical Physics,\\
       Bern University, Switzerland\\
       E-mail: \email{wiese@itp.unibe.ch}}

\abstract{We show that exotic phases arise in generalized lattice gauge theories
known as quantum link models in which classical gauge fields are replaced by 
quantum operators. While these quantum models with discrete variables have a finite-dimensional
Hilbert space per link, the continuous gauge symmetry is still exact. An efficient cluster
algorithm is used to study these exotic phases. The $(2+1)$-d system is confining at zero
temperature with a spontaneously broken translation symmetry. A crystalline phase exhibits 
confinement via multi-stranded strings between charge-anti-charge pairs. A phase transition between 
two distinct confined phases is weakly first order and has an emergent spontaneously 
broken approximate $SO(2)$ global symmetry. The low-energy physics is
described by a $(2+1)$-d $\mathbb{R}P(1)$ effective field theory, perturbed by a dangerously
irrelevant $SO(2)$ breaking operator, which prevents the interpretation of the emergent
pseudo-Goldstone boson as a dual photon. This model is an ideal candidate to be implemented in 
quantum simulators to study phenomena that are not accessible using Monte Carlo simulations 
such as the real-time evolution of the confining string and the 
real-time dynamics of the pseudo-Goldstone boson.}

\FullConference{31st International Symposium on Lattice Field Theory
                - LATTICE 2013\\
		July 29 - August 3, 2013\\
		Mainz, Germany}

\begin{document}

\section{Introduction}
 Lattice gauge theories have made fundamental contributions to our
 understanding of strongly correlated systems. In particle physics, the 
$SU(3)$ lattice gauge theory with Wilson and staggered fermions, and more 
recently with twisted mass, overlap, and domain wall fermions, are extensively used 
to investigate the properties of Quantum Chromodynamics (QCD). In condensed 
matter physics, the study of deconfined quantum critical points, as well as
of spin liquids, employ $U(1)$ lattice gauge theories. The toric code studied in the
condensed matter and the quantum information community is a $\mathbf{Z}(2)$
lattice gauge theory.

 Non-perturbative studies of lattice gauge theories almost exclusively use 
Monte Carlo simulations in Euclidean space-time. 
The conventional formulation is due to Wilson, using parallel transporter matrices 
on the links of the lattice and matter fields on the sites. This method has been
very successful in several respects, such as in the ab-initio calculation of the
hadron spectrum as well as the nature of the finite temperature transition of 
strongly interacting matter in QCD. However, there are certain important problems
where this method fails. Two of the most important examples are the physics at 
non-zero baryon density and the real-time evolution of quantum systems. In these
cases, the weights to be sampled with Monte Carlo become negative, or even
complex. The importance sampling fails, thus leading to a sign problem.

  The rapid development of the field of ultra-cold atoms in optical lattices
suggests a remarkable way out of this problem. The basic idea that the use of 
quantum variables could speed up the simulation tremendously was conceived early on 
\cite{Fey82}. Special purpose quantum computers, known as 
quantum simulators \cite{Cir95}, are used as digital \cite{Llo96} or 
analog devices \cite{Jak98} to simulate strongly coupled quantum systems.
Recently, the use of quantum simulators to study the real-time evolution in 
gauge theories and their phase structure in the context of particle physics
has been proposed 
\cite{Zoh11,Szi11,Liu12,Zoh12,Ban12,Ban13,Zoh13a,Zoh13b,Wie13}. The idea 
behind the quantum simulator constructions  is that the quantum mechanical
nature of quarks and gluons can be embodied by ultra-cold atoms in optical lattices. The 
interactions between the atoms can be tuned, so that they follow a properly 
designed Hamiltonian. The quantum degrees of freedom evolve according to this
Hamiltonian and the sign problem does not arise. A number of important models, such as
the Bose-Hubbard model and the toric code, have already been quantum simulated 
using similar methods \cite{Gre02,Bar11}. 
 
 In this context, the use of alternative formulations of gauge theories is
highly desirable, the principal motivation being the identification of models
with a finite-dimensional Hilbert space at each link or site which can be 
realized with a few quantum states of a cold atom system. The Hamiltonian 
formulation of Wilson's lattice gauge theory has an infinite-dimensional 
Hilbert space at each link due to the use of continuously varying fields. 
Quantum link models (QLMs) provide such an alternative formulation of gauge theories
\cite{Hor81,Orl90,Cha97,Bro99,Bro04} which realize continuous gauge symmetry 
with generalized quantum spins associated with the links of a lattice. 
They constitute an extension of the Wilsonian formulation of lattice gauge theories.
Indeed, in certain limiting cases, Wilson's lattice gauge theories can 
be obtained from quantum link models \cite{Sch00}.
Because they use discrete degrees of freedom, the Hilbert space of quantum link models 
at every link is finite-dimensional in a completely gauge invariant way. 
This enables a direct connection with ultra-cold atoms in optical lattices,
where the generalized quantum spins can be represented by these atoms.
Therefore, quantum link models are ideal candidates to be implemented
in cold atom systems.

 The regularization of a $d$-dimensional quantum field theory formulated with 
discrete quantum variables in $(d+1)$-dimensions (instead of using classical fields) is
 known as the D-theory formulation \cite{Bro04}. The classical fields in the 
$d$-dimensional quantum field theory emerge as low-energy effective degrees of
 freedom of the discrete variables when the $(d+1)$-dimensional theory has a 
massless Coulomb phase. When the extra Euclidean dimension is made small in
units of the correlation length, a $d$-dimensional theory emerges by
dimensional reduction. For example, in the D-theory formulation of QCD, the 
confining gluon field emerges by dimensional reduction from a deconfined Coulomb
 phase of a $(4+1)$-d $SU(3)$ quantum link model. Chiral quarks can be included
naturally as domain wall fermions located at the two 4-d sides of a $(4+1)$-d 
slab \cite{Bro99}.

  Quantum link models also provide a platform for developing efficient 
simulation algorithms. As they are formulated with discrete variables, they are
 natural candidates to develop cluster algorithms. The phase diagrams of these
models are obviously interesting to study. Since they are generalized 
lattice gauge theories, new phases arise, which have not been observed in 
Wilson-type lattice gauge theories. Also, once quantum simulators are being built, 
they need to be validated against controlled 
classical computations. By developing methods to simulate quantum link 
models, static quantities can be calculated to benchmark the quantum 
simulators. Finally, methods developed for simulating link models might be 
applicable to solve some of the sign problems in traditional Wilson-type 
theories at non-zero density.

 In this article, we report on a study of the $(2+1)$-dimensional $U(1)$ quantum link
model and show that, despite its structural simplicity, it has a very rich
phase diagram \cite{Ban13a}. This model has exotic confining 
phases where the confining string joining a static charge-anti-charge pair 
splits into distinct fractionalized flux $\frac{1}{2}$ strands. There are two 
such phases, each characterized by a distinct pattern of discrete symmetry 
breaking, separated by a weak first-order transition. Around the phase 
transition point, there is a spontaneously broken approximate global $SO(2)$
symmetry arising dynamically. The resulting pseudo-Goldstone boson can be 
described via an effective field theory. When realized 
with quantum simulators, this model would be able to demonstrate the power of 
gauge theory simulators by quantum simulating the dynamics of the confining 
string and the pseudo-Goldstone boson. 
%Work is under way to develop a scheme 
%for simulating this model using superconducting Qubits \cite{MarXX}.

\section{The $(2+1)$-d Quantum Link Model}
In the Wilson formulation of $U(1)$ lattice gauge theory, the Hamiltonian takes the
form
\begin{gather}
 H = \frac{g^2}{2} \sum_{x,i} e^2_{x,i} - \frac{1}{2g^2}\sum_{\square} ( u_{\square}
+ u_{\square}^{\dagger} ),
\end{gather}
where the second sum is over all plaquettes and the plaquette variables are 
$u_{\square} = u_{x,i}u_{x+\hat{i},j}u_{x+\hat{j},i}^{\dagger}u_{x,j}^{\dagger}$,
$u_{x,i} = \exp(i \varphi_{x,i}) \in U(1)$. 
In this formulation they are operators acting in an infinite-dimensional
Hilbert space for each link. The electric field operator $e_{x,i}$ describes
the kinematics of $u_{x,i}$, 
\begin{gather}
e_{x,i} = -i \partial_{\varphi_{x,i}}, \qquad [e_{x,i},u_{y,j}] = u_{x,i} \delta_{xy}\delta_{ij}, \qquad
[e_{x,i},u_{y,j}^{\dagger}] = -u_{x,i}^{\dagger}\delta_{xy}\delta_{ij}, \qquad [u_{x,i},u_{y,j}^{\dagger}]
= 0.
\end{gather}
The Hamiltonian is invariant under gauge transformations since it commutes
with their generators $G_x = \sum_i (e_{x,i} - e_{x-\hat{i},i})$. A general gauge
transformation takes the form $u_{x,i}' = e^{i\alpha_x} u_{x,i}
e^{-i\alpha_{x+\hat{i}}}$.

The $(2+1)$-d $U(1)$ quantum link model is formulated in a similar way. Its
Hamiltonian can be written as
\begin{gather}
 H = \frac{g^2}{2} \sum_{x,i} E^2_{x,i} -J \sum_{\square} \left[ U_{\square} +
U_{\square}^{\dagger} - \lambda \left( U_{\square} + U_{\square}^{\dagger}
\right)^2\right]
\label{H_QLM}
\end{gather}
where the sum is over all plaquettes and the plaquette variables are defined
in terms of quantum link operators, 
$U_{\square} := U_{x,i}U_{x+\hat{i},j}U_{x+\hat{j},i}^{\dagger}U_{x,j}^{\dagger}$. 
In contrast to Wilson's lattice gauge theory, the operators in the $U(1)$ QLM are
given by a finite-dimensional representation of the embedding algebra $SU(2)$,
thus leading to a finite-dimensional Hilbert space per link \cite{Cha97}.
The quantum link variables are quantum spin raising operators 
$U_{x,i} = S_{x,i}^1 + i S_{x,i}^2 = S^+_{x,i}$ for the electric fluxes $E_{x,i} =
S^3_{x,i}$, 
while the operators $U_{x,i}^{\dagger}$ are flux lowering operators $S_{x,i}^-$. 
The operators $U_{x,i}$, $U_{x,i}^{\dagger}$ and $E_{x,i}$ obey the same commutation 
relations as their counterparts in Wilson's lattice gauge theory, except 
that the quantum link operators do not commute with their adjoint, i.e. 
$[U_{x,i}, U_{x,i}^{\dagger}] = 2\, E_{x,i}$. The Hamiltonian in (\ref{H_QLM}) 
is again gauge invariant as it commutes with the generators of infinitesimal $U(1)$ 
gauge transformations,
\begin{gather}
 G_x = \sum_i \left( E_{x,i} - E_{x-\hat{i},i}\right).
\end{gather}
The link operators transform as $U_{x,i}' = e^{i\alpha_x} U_{x,i}
e^{-i\alpha_{x+\hat{i}}}$ under gauge transformations. 
Physical states $|\psi\rangle$ again have to be gauge invariant,
i.e.\ they obey the Gauss law $G_x |\psi\rangle = Q_x | \psi\rangle$, where $Q_x$
is zero unless one places a static charge at the point $x$.

\begin{figure}[ht]
\begin{center}
 \includegraphics[width=9cm]{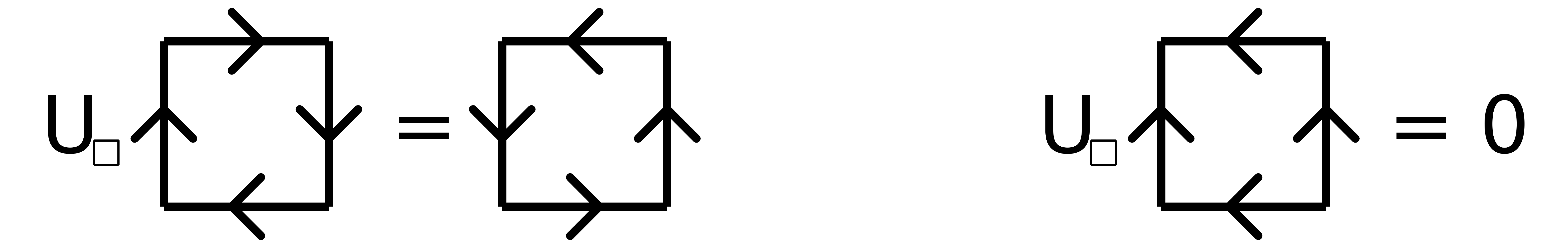}
 \caption{Action of the plaquette operator on flippable plaquettes (left) and 
non-flippable plaquettes (right). The arrows indicate the direction of the electric 
flux $E_{x,i} = \pm \frac{1}{2}$.}
 \label{ActionOfH}
\end{center}
\end{figure}

In this work we consider the simplest possible representation for the quantum links,
namely the spin $\frac{1}{2}$ representation. This leads to a 2-dimensional
Hilbert space per link, thereby ensuring the feasibility of exact 
diagonalization studies, as will be explained below. The $E_{x,i}^2$ term then becomes a 
trivial additive constant and is therefore omitted in the following. The second
term in the Hamiltonian (\ref{H_QLM}) flips loops of electric fluxes flowing 
around elementary plaquettes and annihilates non-flippable plaquettes 
as depicted in Fig.\ \ref{ActionOfH}. In the spin $\frac{1}{2}$ representation,
the third term, proportional to $\lambda$, is known as the 
Rokhsar-Kivelson (RK) term which counts flippable plaquettes. This can be seen by
noting that $U_{\square}^2 = \left(U_{\square}^{\dagger}\right)^2 = 0$, since a
single spin $\frac{1}{2}$ 
cannot be raised more than once. The remaining terms are of the form 
$U_{\square} U_{\square}^{\dagger}$, i.e. they project onto the subspace of 
flippable plaquettes. 

\begin{figure}[htb]
\begin{center}
 \includegraphics[width=10cm]{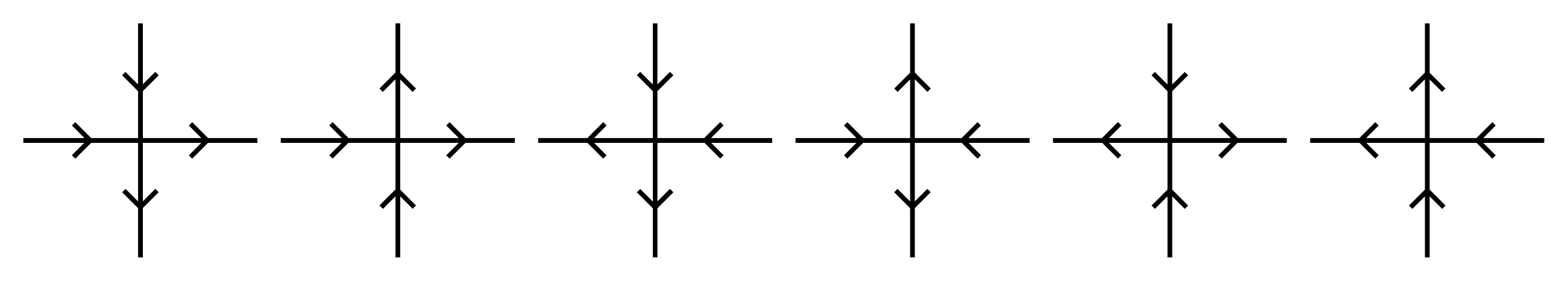}
 \caption{All gauge invariant configurations of fluxes at a site $x$.}
 \label{GaussLaw}
\end{center}
\end{figure}

The Gauss law together with the finite-dimensional link Hilbert space
reduces the number of allowed states per site from $2^4 = 16$ down to the
6 configurations shown in Fig.\ \ref{GaussLaw}. Adding static charges reduces this
number even further. Without this reduction, it would not be practical to apply
exact diagonalization methods on reasonably large lattices.

The Hamiltonian respects the usual geometric symmetries of the
square lattice, e.g. it is invariant under translations by multiples
of the lattice spacing $a$ and under 90 degrees rotations. For our purposes, 
it suffices to consider translations T only. The lattice translation invariance
characterizes each energy eigenstate by its lattice momentum 
$p = (p_1,p_2) \in (-\pi,\pi]^2$. Additionally, the charge conjugation C
symmetry is also present. It replaces $U_{x,i}$ by $U_{x,i}^{\dagger}$
and reverses all electric fluxes, i.e. $E_{x,i}$ goes to $-E_{x,i}$. 
The associated quantum number is the charge conjugation parity $C = \pm$.
Another important global symmetry is the $U(1)$ center symmetry on periodic
lattices associated with ``large'' gauge transformations.
These are given by transformations that commute with the Hamiltonian
but cannot be expressed through ``small'' periodic gauge transformations.
On an $L_1 \times L_2$ lattice they are generated by
\begin{gather}
 E_i = \frac{1}{L_i} \sum_x E_{x,i}.
\end{gather}

\begin{figure}[htb]
\begin{center}
\includegraphics[scale=0.75]{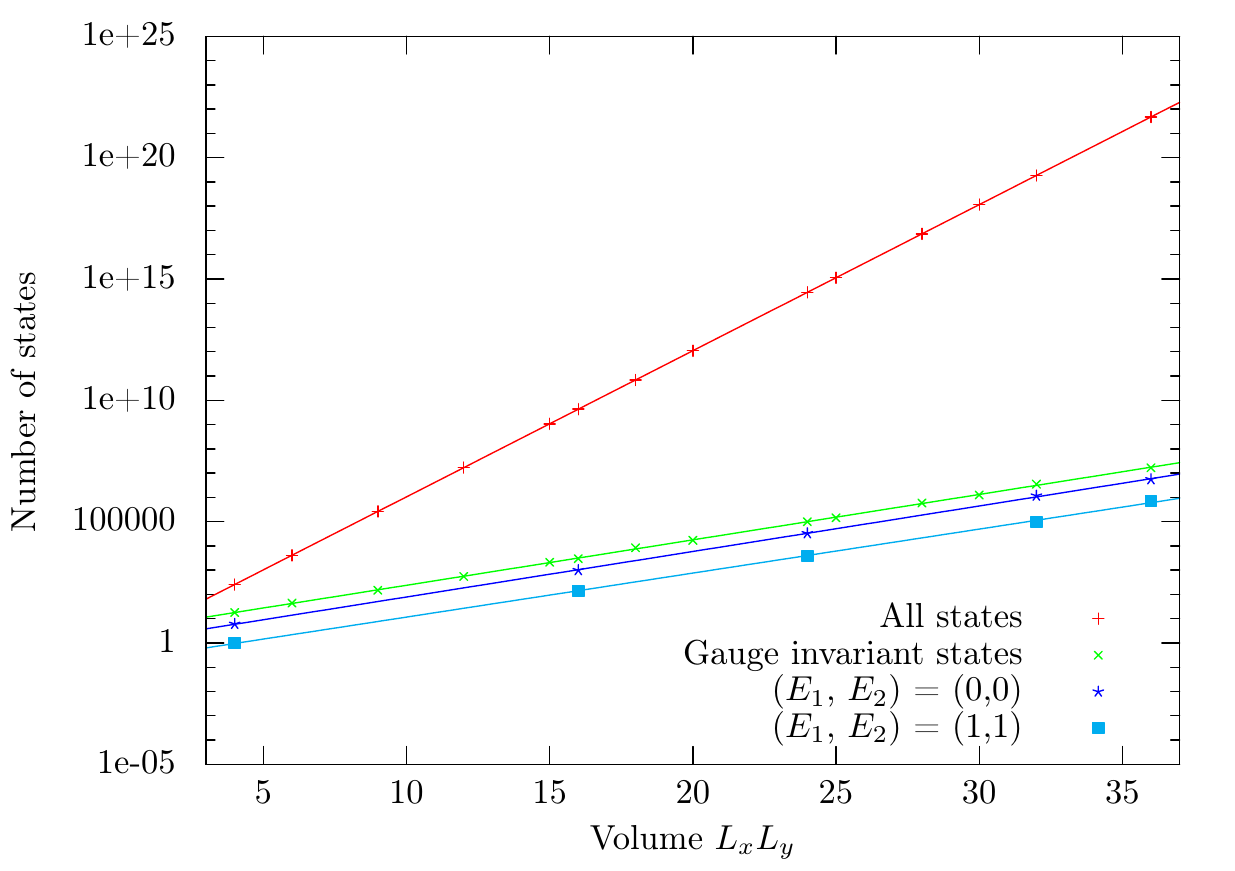}
\end{center}
\caption{The total number of states in the $(2+1)$-d $U(1)$ quantum link model,
both with and without imposing Gauss' law (denoted by green "$\times$" and red "+"
respectively). Even though the Gauss law constraint drastically decreases the 
number of states, the rise in the
number of states is still exponential in the volume. The number of states in individual
flux sectors are shown by blue stars and filled squares. Only lattices with an
even extent allow the flux sectors $(0,0)$ and $(1,1)$.}
\label{states}
\end{figure}

\section{Exact Diagonalization and Cluster Algorithm Tools}
 We have studied the model using both Exact Diagonalization (ED) and Quantum
Monte Carlo simulations (QMC).  Exact Diagonalization studies were performed on lattices with spatial 
extents $4 \times 4$, $4 \times 6$ and $6 \times 6$. These systems comprise of 32, 48, and 72 
quantum link spins, respectively. While the sizes may seem
small, this already competes with the largest spin systems that have been 
subjected to ED on PC clusters. Naively, this would have implied Hilbert
spaces with $2^{32}$, $2^{48}$, and $2^{72}$ states, respectively. 
The Gauss law constraint, however, reduces the number of states considerably,
which makes the study of systems with as many as 72 spins feasible. 
Fig.\ \ref{states} shows the number of states as a function of the volume 
with and without applying Gauss' law. In these studies, the Hamiltonian was
separately diagonalized in each flux winding number sector, 
thereby reducing the Hilbert space further than with imposing only the Gauss
law constraint.

\begin{figure}
\begin{center}
\includegraphics[scale=0.55]{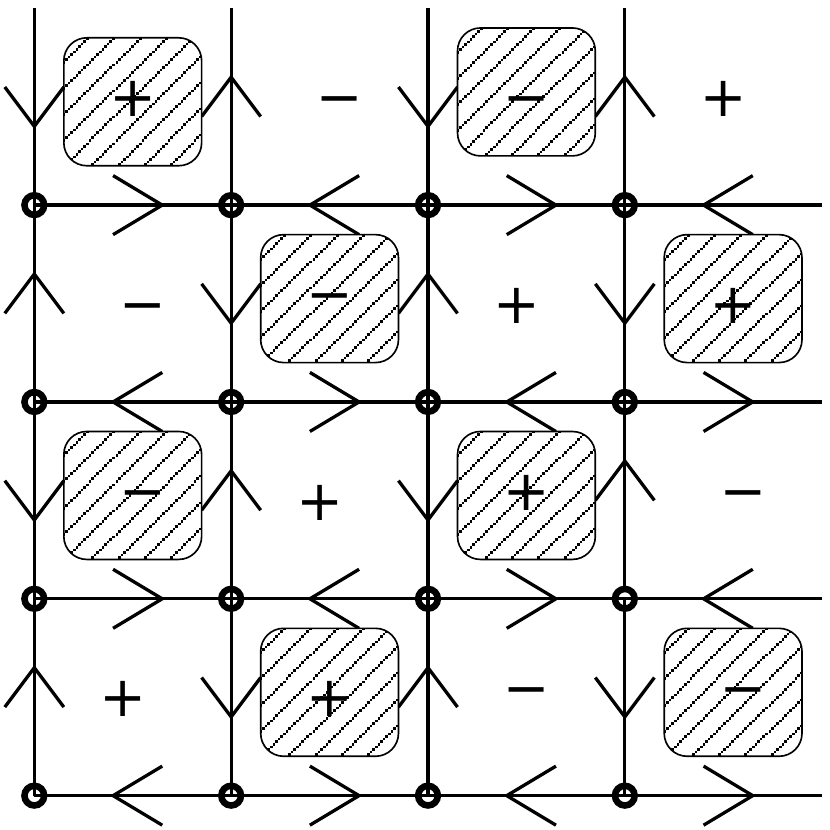}
\hspace{1.2cm}
\includegraphics[scale=0.35]{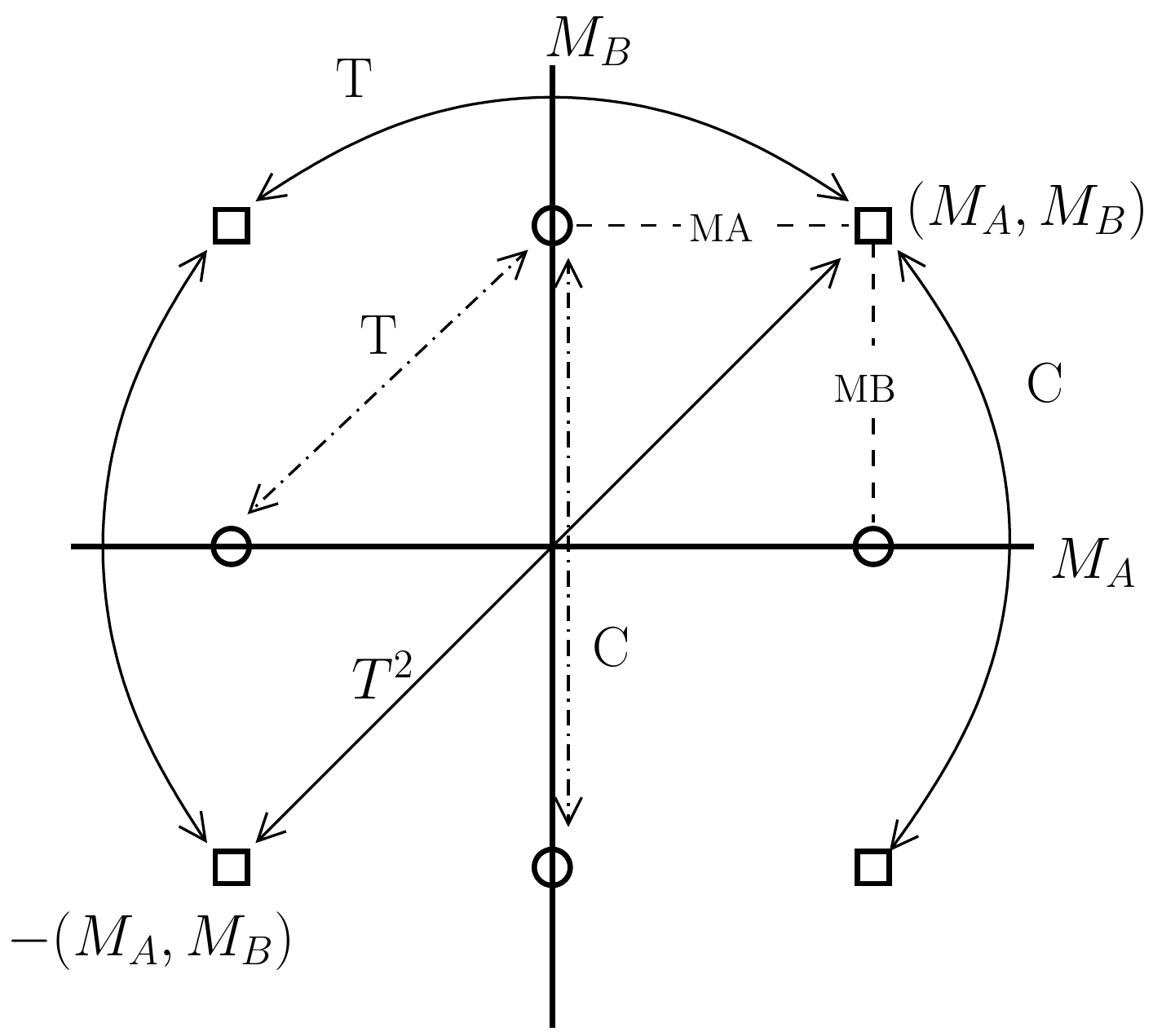}
\end{center}
\caption{(a) Mapping of an electric flux configuration (shown with arrows on
the links) to a height configuration (shown with $+$ and $-$ variables at the 
centers of the plaquettes). Every time one crosses a flux pointing right or upwards, 
the orientation of the plaquette variable is changed, while it remains
 unchanged if a left or downward pointing flux is crossed. (b) The effect
of the symmetry transformations C and T on the two-component order parameter 
$(M_A,M_B)$. The former is equivalent to a reflection on the $M_A$ axis, while the
latter is an anti-clockwise rotation by $\frac{\pi}{2}$. Performing T twice is
equivalent to rotating by $\pi$, and leads back to the starting 
configuration, since $-(M_A,M_B)$ is an equivalent "gauge" copy of $(M_A,M_B)$.}
\label{OP}
\end{figure}

 We also developed an efficient cluster algorithm to simulate the model in the 
dual representation. A duality transformation can be used to transform the 
Hamiltonian of the $(2+1)$-d quantum link model 
into that of a $(2+1)$-d quantum height model. This transformation is an 
exact rewriting of the partition function in terms of new degrees of freedom, 
which are quantum $\mathbb{Z}(2)$ variables located at the centers of
the plaquettes. As shown in Fig.\ \ref{OP}(a), every flux configuration 
can be mapped to a height configuration. A configuration of quantum height 
variables 
\begin{equation}
h^A_{\widetilde x} = 0,1;~~h^B_{\widetilde x} = \pm \frac{1}{2},
\end{equation}
located at the dual sites $\widetilde x = (x_1 + \frac{1}{2},x_2 + \frac{1}{2})$, 
is associated with a flux configuration 
\begin{equation}
E_{x,x+\hat i} = [h^X_{\widetilde x} - h^{X'}_{\widetilde x+ \hat i - \hat 1 - \hat 2}] 
\mbox{mod} 2 = \pm \frac{1}{2};~~X,X' \in \{A,B\}.
\end{equation}
The cluster algorithm is then constructed by dividing the lattice into
 two sublattices $A$ and $B$ (illustrated by shaded and unshaded squares in Fig.\ 
\ref{OP} (a)). 
The $U(1)$ Gauss law constraint is implemented in the cluster building rules, 
which ensure that only the configurations with net zero charge at the 
vertices are generated. The details of the dualization procedure as well as the
algorithm will be presented elsewhere \cite{BanXX}. 

 We define a 2-component order parameter $(M_A,M_B)$, associated with the 
even and odd sublattices $A$ and $B$, to characterize the different phases of 
the model. These distinguish the two different symmetry breaking patterns
we encountered in our study. In terms of the height variables 
associated with the center of the plaquettes, they are defined as
\begin{equation}
M_X =  \sum_{\widetilde x \in X} s^X_{\widetilde x} h^X_{\widetilde x};
~~ \textrm{where}~~ s^A_{\widetilde x} = (-1)^{(\widetilde x_1 - \widetilde x_2)/2} 
~~ \textrm{and}~~ s^B_{\widetilde x} = (-1)^{(\widetilde x_1 - \widetilde x_2 + 1)/2}.
\end{equation}
 Under C and T they transform as $^C M_A = M_A$, 
$^C M_B = - M_B$, $^T M_A = - M_B
$, $^T M_B = M_A$. It should be pointed out that $\pm (M_A,M_B)$ represents the 
same physical configuration because shifting the height variables to 
$h^X_{\widetilde x}(t)' = [h^X_{\widetilde x}(t) + 1] \mbox{mod} 2$ leaves 
the electric flux configuration unchanged. The various transformations are  
illustrated in Fig.\ \ref{OP}(b).
\begin{figure}
\begin{center}
\includegraphics[scale=0.4]{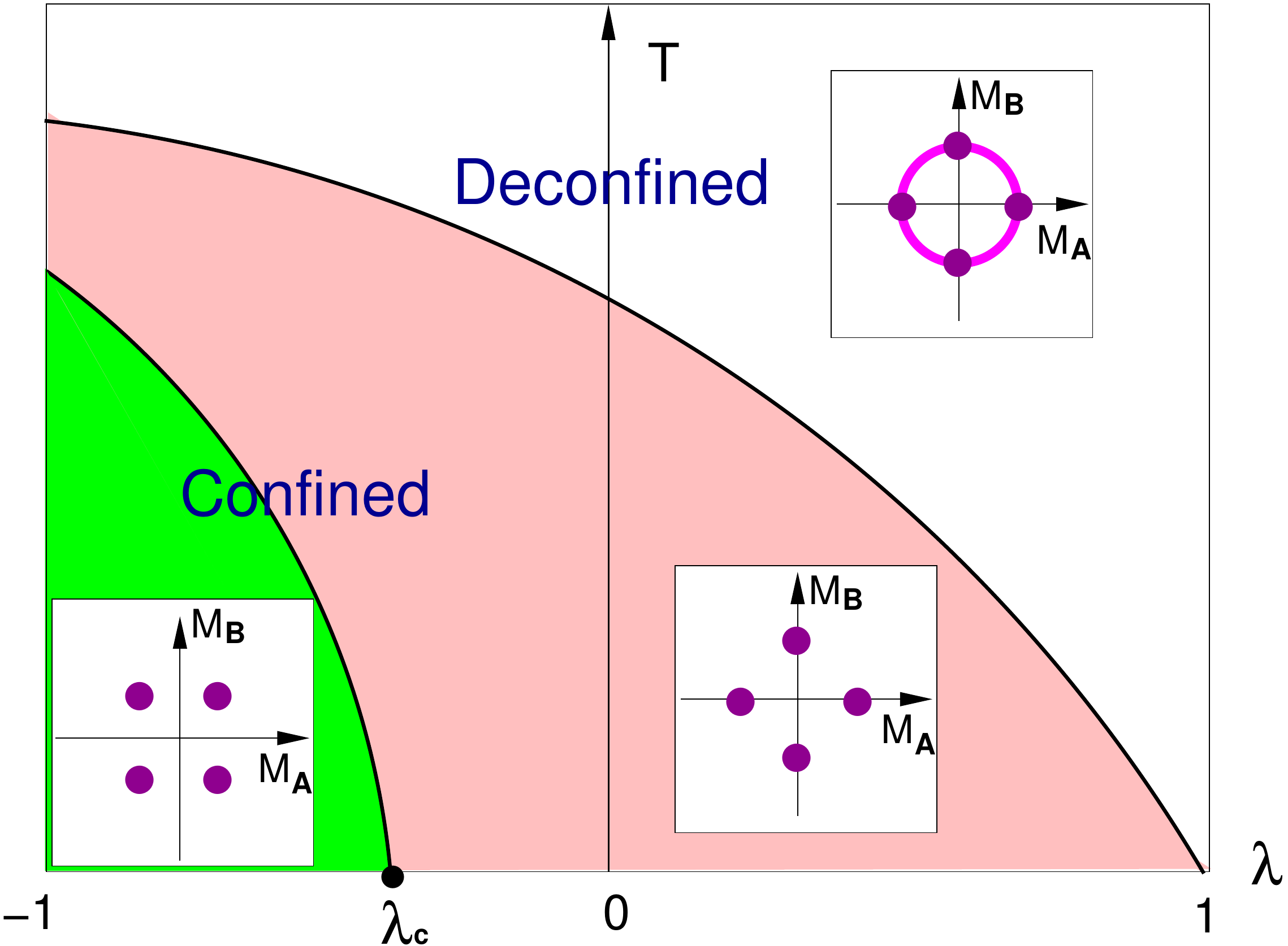}
\end{center}
\caption{Schematic sketch of the $\lambda-T$ phase diagram. The insets
indicate the location of the peaks in the probability distribution of 
the order parameter $p(M_A,M_B)$.}
\label{phdiag}
\end{figure}

\section{Phase Diagram, Order Parameters and Confining Strings}
\begin{figure}[htb]
\begin{center}
\includegraphics[scale=0.6]{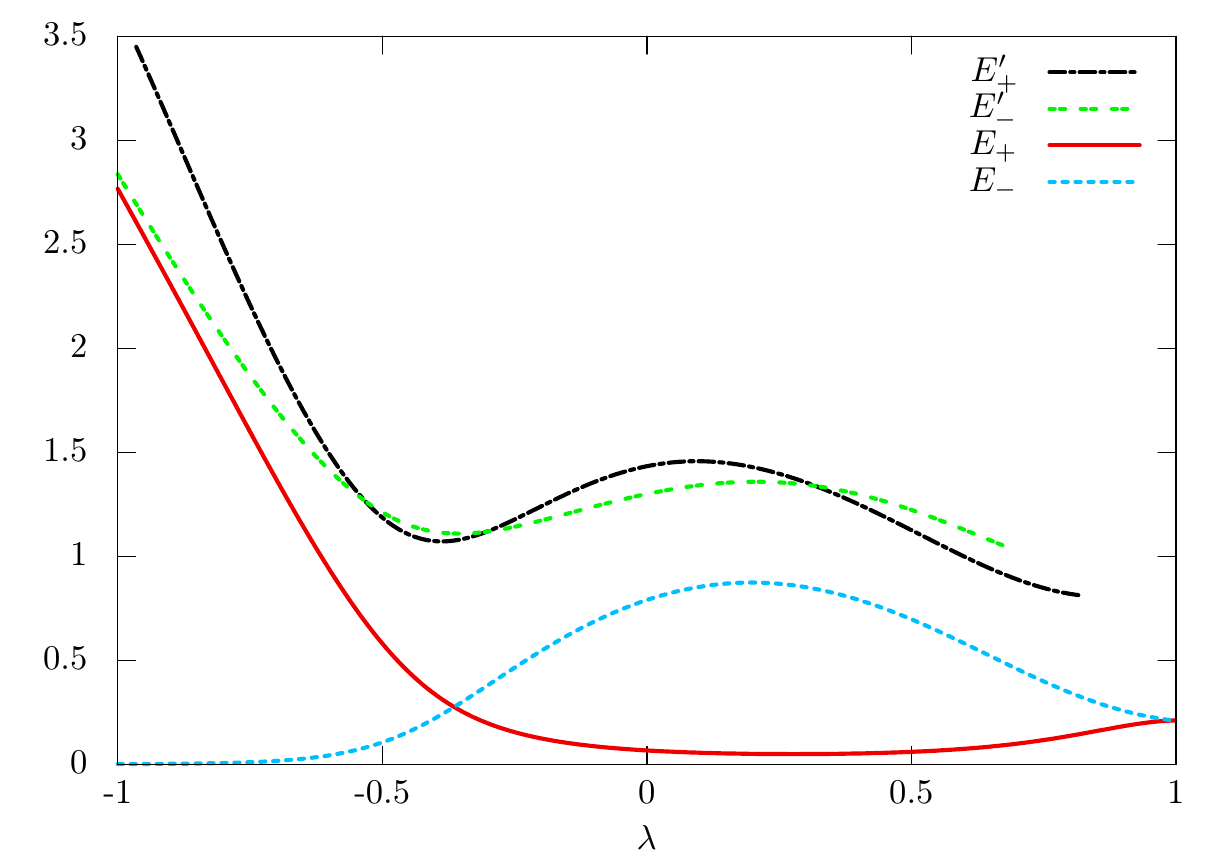}
\end{center}
\caption{Energy gaps of the four lowest states on the $(6 \times 6)$ lattice as
 a function of $\lambda$ relative to the ground state energy. The states with energies
$E_+$ and $E_-$ both have non-zero momenta $(\pi,\pi)$, and are degenerate with the ground state 
(which has zero momentum and positive C parity) in the infinite volume limit. 
The state with energy $E_-$ has negative C parity and the state with energy $E_+$ 
has positive C parity. This implies that the spontaneous symmetry breaking pattern 
changes as one crosses $\lambda_c$. The next higher states have energies $E^\prime_\pm$ 
and carry momentum $(0,0)$. They have C parity $\pm$ and are used 
to determine the parameters of the effective theory, as described in Section 5.}
\label{statecross}
\end{figure}

 The phase diagram of the model, shown in Fig.\ \ref{phdiag}, was studied as 
a function of both $\lambda/J$ and $T/J$, where $T$ is the temperature. 
For convenience, we work with 
units in which $J=1$. At zero temperature, the model has two phases 
characterized by different symmetry breaking patterns for C and T. For large 
negative $\lambda$, both C and T are spontaneously broken. As $\lambda$ is 
increased beyond a critical value $\lambda_c$, the model undergoes a weak 
first order phase transition into a phase where T, but not C, 
is spontaneously broken.
\begin{figure}[htb]
\begin{center}
\includegraphics[scale=0.4]{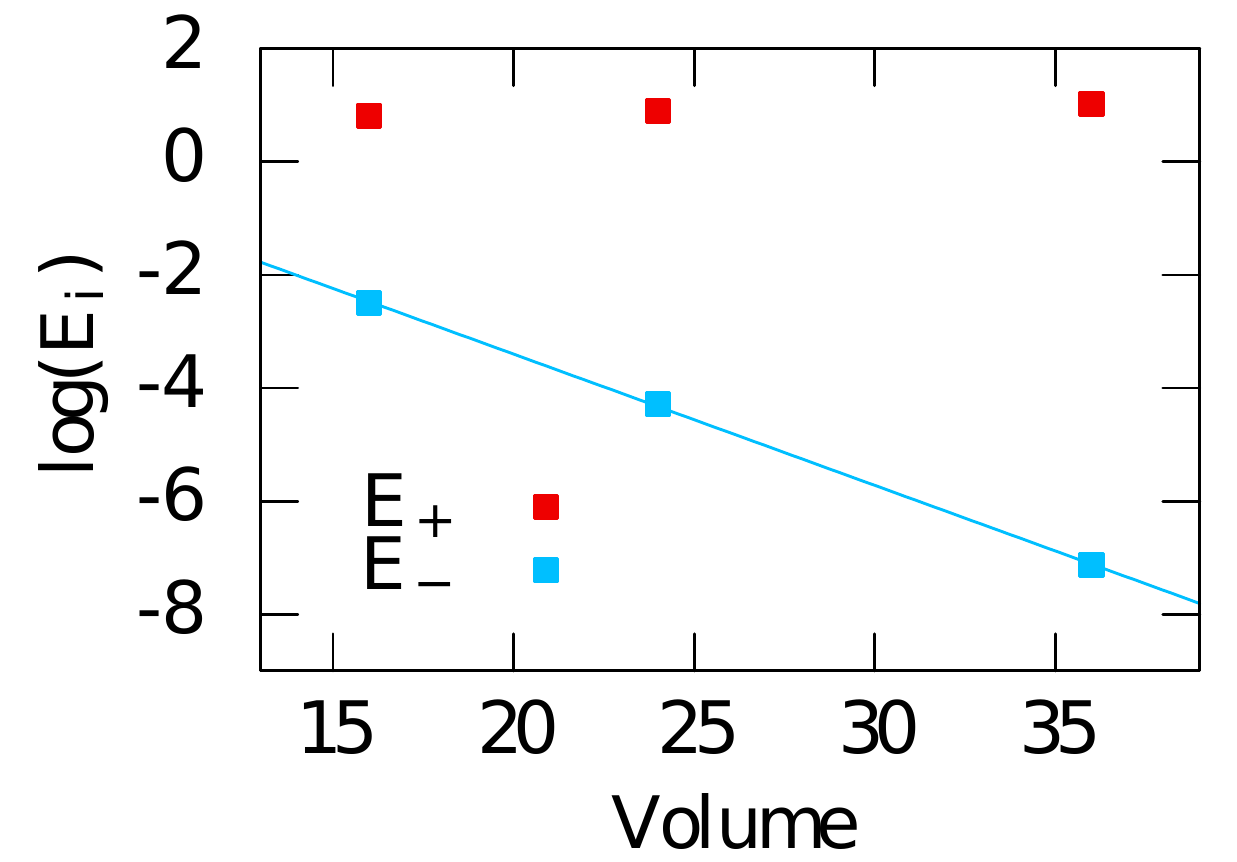}
\hspace{1.2cm}
\includegraphics[scale=0.4]{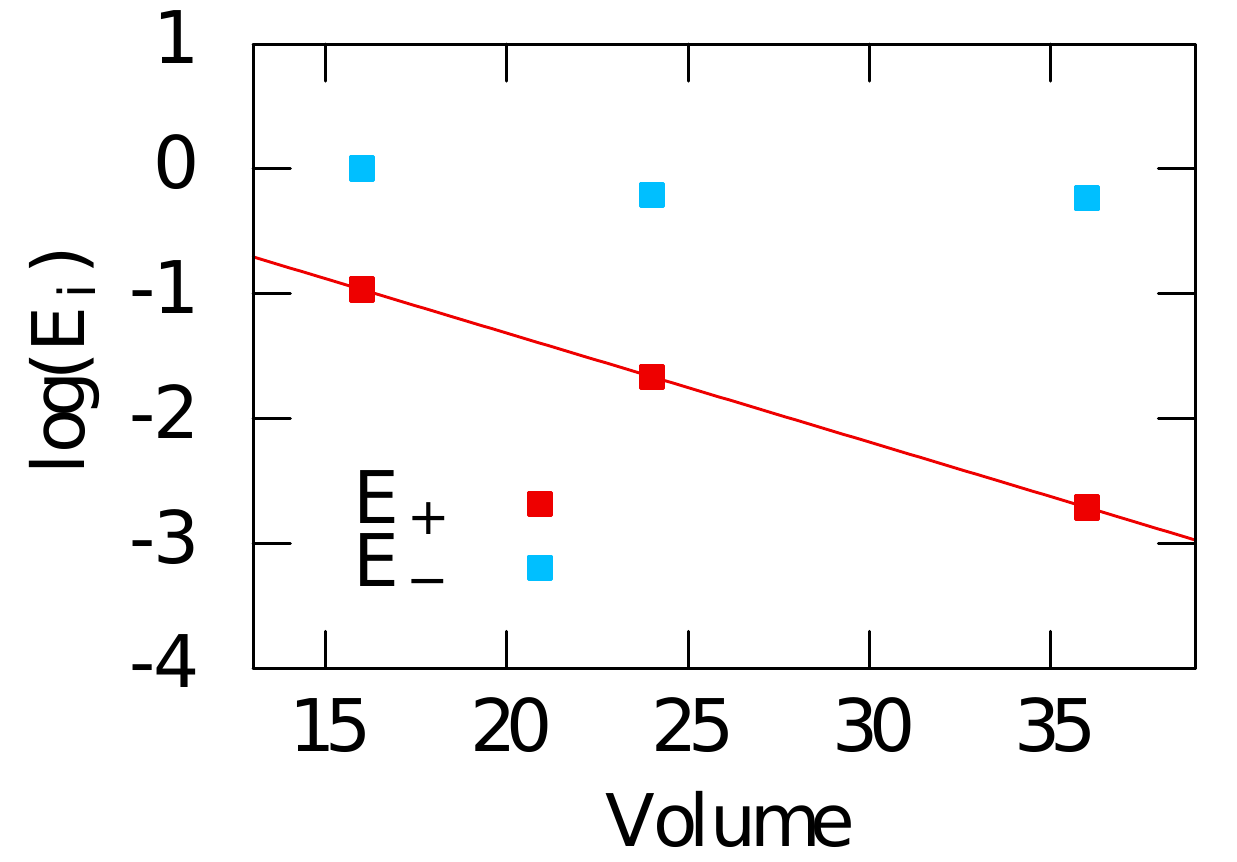}
\end{center}
\caption{Finite-volume energy spectrum of the two lowest excitations above the ground 
state. At $\lambda~=~-1$, the state with quantum numbers $p=(\pi,\pi)$ and 
$C=-$ degenerates with the ground state (left). At $\lambda~=~0$, the state with 
quantum numbers $p=(\pi,\pi)$ and $C=+$ degenerates with the ground state (right). The 
ground state has positive C parity and zero momentum.}
\label{finV}
\end{figure}

 It is interesting to note that the ED results provide important insights into
the phase diagram. Fig.\ \ref{statecross} shows the energy gaps of the four
lowest energy states. For $\lambda < 1$, the ground state has momentum $(0,0)$ and
 is even under charge conjugation (i.e. $C~=~+$). For $\lambda < \lambda_c$, 
the first excited state has quantum numbers $C = -$, $p = (\pi,\pi)$. Its energy
 gap to the ground state, $E_- \sim \exp(- \sigma_- L_1 L_2)$, decreases 
exponentially with the volume $L_1 L_2$, thus indicating the spontaneous 
breakdown of charge conjugation C and the translation T by one lattice spacing 
(in either direction). For $\lambda > \lambda_c$, another state 
$|C=+,p = (\pi,\pi)\rangle$ degenerates with the ground state in the infinite volume
limit, i.e.\ 
$E_+ \sim \exp(- \sigma_+ L_1 L_2)$, indicating that C is now restored, while T 
remains spontaneously broken. The finite-volume scaling of the spectrum 
indicating these symmetry breaking patterns is shown in Fig.\ \ref{finV}.

\begin{figure}
\begin{center}
\includegraphics[scale=0.4]{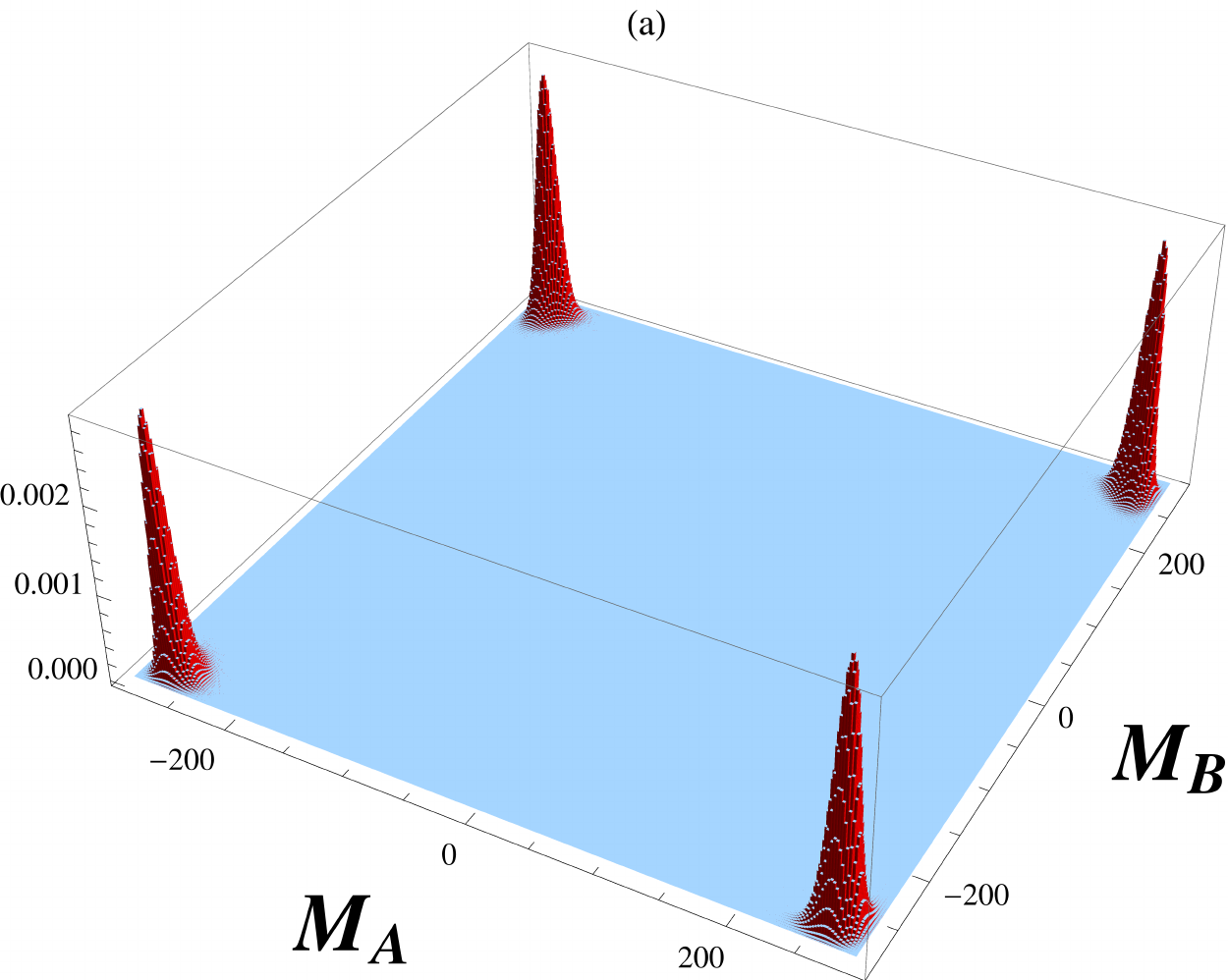}
\hspace{1.2cm}
\includegraphics[scale=0.4]{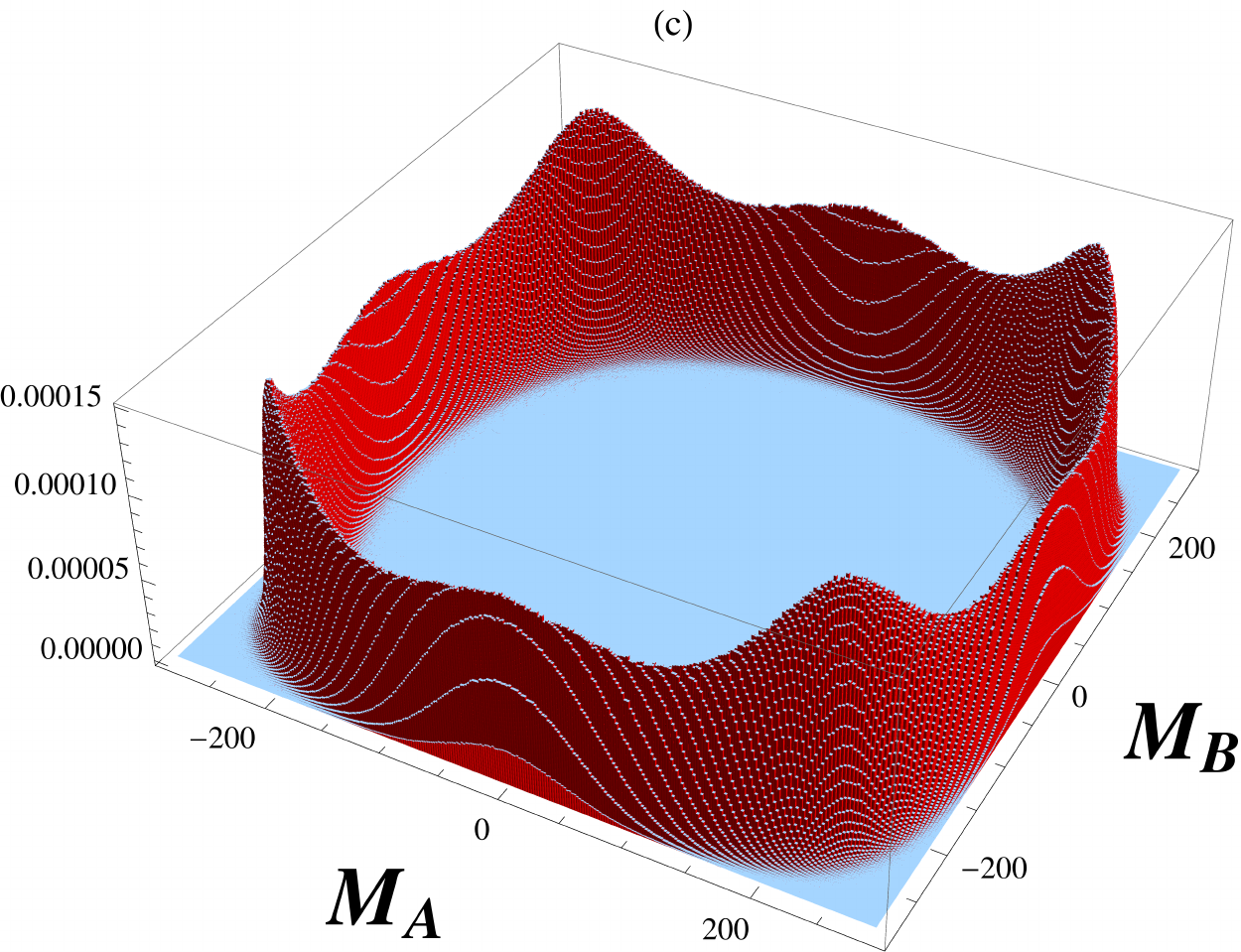}
\includegraphics[scale=0.4]{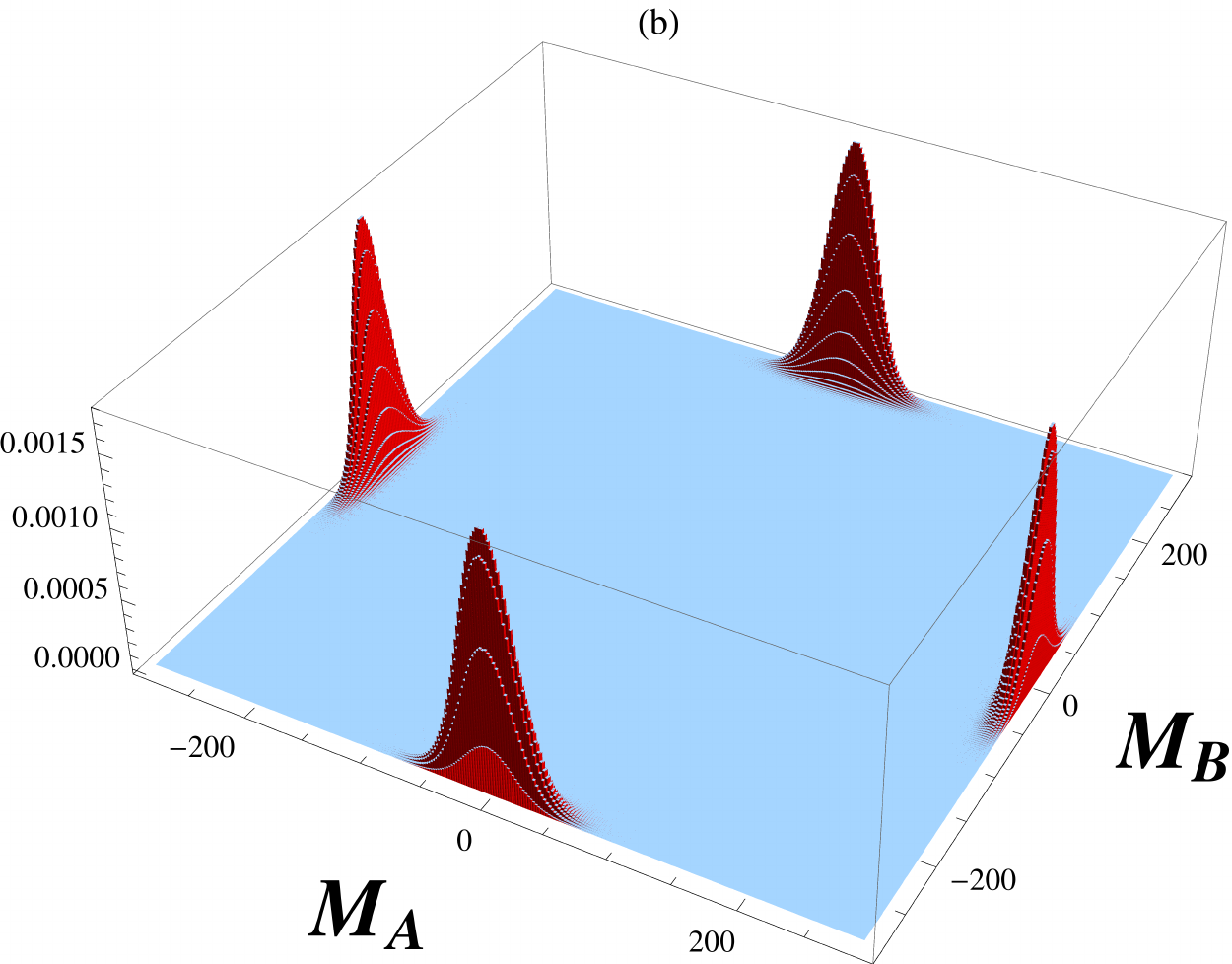}
\hspace{1.2cm}
\includegraphics[scale=0.4]{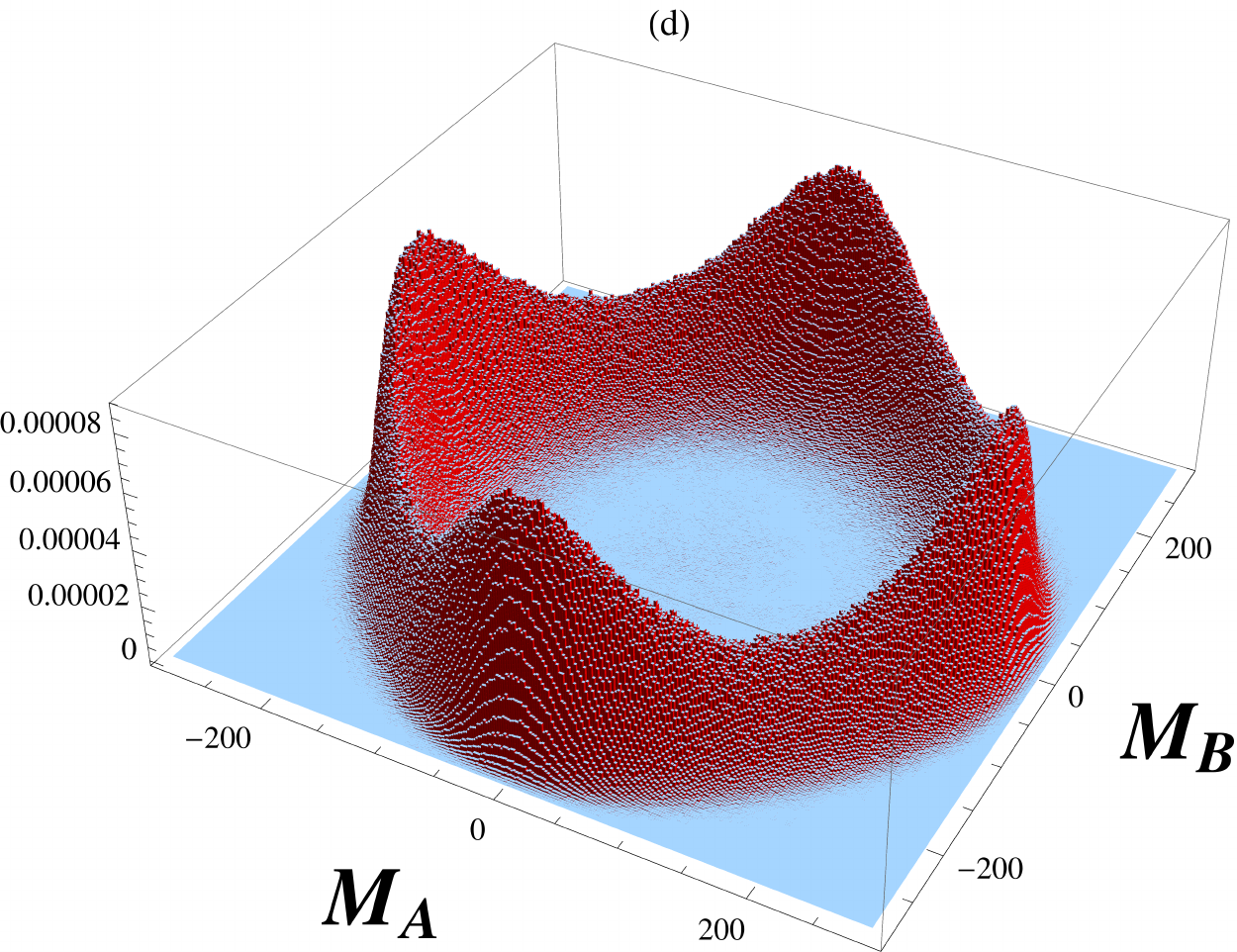}
\end{center}
\caption{Distribution of the order parameter $(M_A,M_B)$ at $\lambda = -1$, 
$\lambda_c$, and 0 at $T = 0$ (a,b,c), and at $\lambda = 0$, $T>T_c$ (d).}
\label{probdist}
\end{figure}

 These symmetry breaking patterns are clearly distinguished by the 
two-component order parameter $M=(M_A,M_B)$. The probability distribution $p(M)$ 
has been measured very accurately 
with the cluster algorithm and is shown in Fig.\ \ref{probdist}
for the different cases. At $\lambda=-1$, 
both sublattices are ordered, giving rise to peaks at the corners of the 
two-dimensional order parameter plane. At $\lambda=0$, only one sublattice is ordered,
which exhibits peaks on the axes. At $\lambda=\lambda_c$, there is an emergent approximate 
global $SO(2)$ symmetry, which manifests itself by an order parameter 
distribution that is nearly circular. There is an
emergent pseudo-Goldstone boson which can be described in terms of a low-energy
effective theory. This is a remarkable phenomenon which mimics some  
features of \emph{deconfined quantum criticality}, widely discussed in the
condensed matter literature \cite{Sen04,Sen04a,San07,Mel08,Jia08,Che13,Tan13,
Dam13,Alb11,Zhu12,Gan13}. At $\lambda=1$ the model reaches its Rokhsar-Kivelson point.
At this point electric flux condenses in the vacuum and the theory deconfines already
at zero temperature.

\begin{wrapfigure}{l}{7.5cm}
\vspace{-0.8cm}
\begin{center}
\includegraphics[scale=0.8]{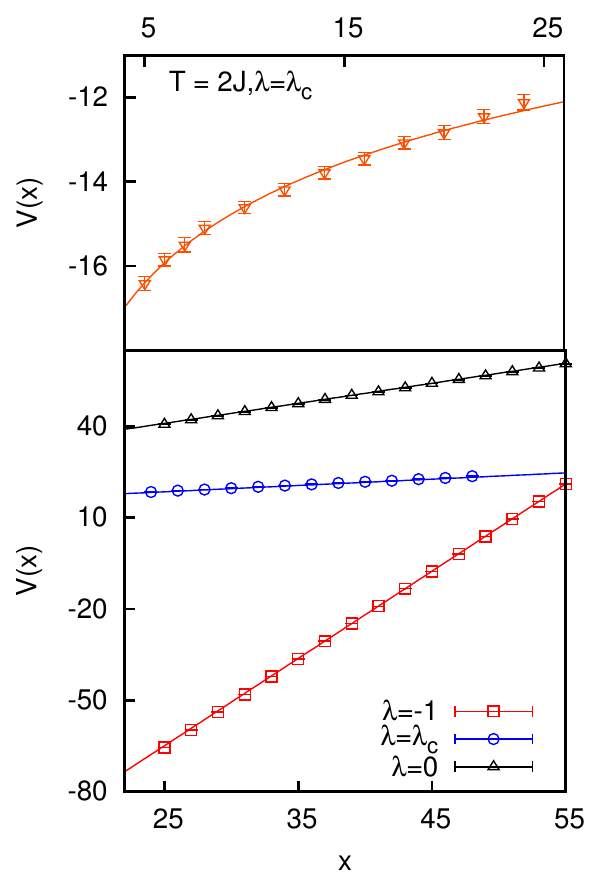}
\end{center}
\vspace{-0.5cm}
\caption{The potential between two static charges $\pm 2$ separated by the 
distance $(x,x)$ along a lattice diagonal, for $\lambda=-1,\lambda_c$, and 0, at
 $T=0$, and at $\lambda = \lambda_c$ for $T=2J$.}
\label{fluxpot}
\vspace{-1.0cm}
\end{wrapfigure}

The phase diagram can also be studied as a function of temperature.
Based on universality arguments, we expect that, the
system undergoes a Berezinski-Kosterlitz-Thouless transition into a deconfined 
Coulomb phase above a temperature $T_c$. However, translation invariance 
still remains broken as evidenced by the corresponding order parameter 
distribution shown in Fig.\ \ref{probdist}(d). At very high temperatures, we 
expect all breaking of translation invariance to disappear.

 Since the quantum link model is a gauge theory in $(2+1)$-dimension, we expect that it 
is linearly confining  for $\lambda < 1$ and $T=0$ \cite{Pol75,Goe81}. A 
standard way of demonstrating this is to place a static charge-anti-charge pair 
at a certain distance $r$, and then study the static potential $V(r)$ as a
 function of $r$. A linearly increasing potential is an unambiguous sign for 
confinement. The string tension $\sigma$ is given by the slope of the static 
potential at large distances. We have studied this by placing 
static charges $Q = \pm 2$ along the lattice diagonal. Our results shown in Fig.\ 
\ref{fluxpot} exhibit linear confinement at large distances, even at the 
phase transition, albeit with a small string tension $\sigma_2 = 0.156(14) J/a $
(compared to $\sigma_2 = 1.97(1) J/a$ at $\lambda=-1$). 
Since we insert the charges explicitly in the simulation, 
our results for the static potential do not suffer from an 
exponentially small signal-to-noise ratio at larger charge-anti-charge separations.

\begin{figure}
\begin{center}
\includegraphics[scale=0.9]{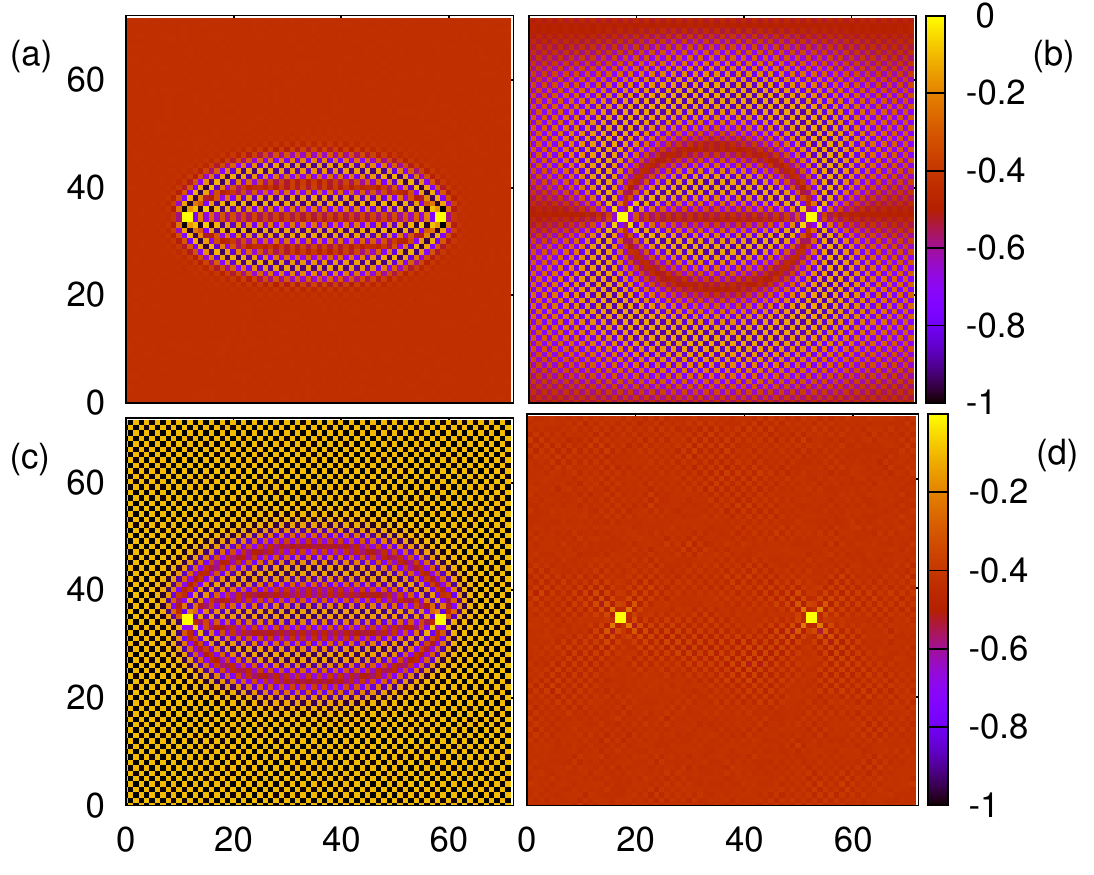}
\end{center}
\caption{Energy density 
$-J \langle U_\Box + U_\Box^\dagger\rangle$ in the presence of two charges $\pm 2$
 for $\lambda = - 1$ (a), $\lambda = \lambda_c$ (b), and $\lambda = 0$ (c) at 
$T = 0$, as well as for $\lambda = 0$ at $T > T_c$ (d).}
\label{halfstrands}
\end{figure}

 Since translation invariance by a single lattice spacing is spontaneously broken in both 
the phases at $\lambda < \lambda_c$ and at $\lambda > \lambda_c$, the resulting 
confined phases are crystalline. The energy density 
$-J \langle U_\Box + U_\Box^\dagger\rangle$ in the presence of two charges $\pm 2$ 
illustrates the nature of the bulk phases. The flux string connecting the 
charges, shown in Fig.\ \ref{halfstrands}, separates into four strands of 
flux $\frac{1}{2}$ that repel each other. The interior of the strands 
consists of the phase that is stable on the other side of the transition. Near 
$\lambda_c$ the flux string undergoes topology change by wrapping one strand 
over the periodic boundary and materializing an additional strand at the edge of
 the system, whose interior then expands to become the new bulk phase (cf. Fig.\
 \ref{halfstrands}(b)). Viewed as interfaces separating bulk phases, the strands 
display the universal phenomenon of complete wetting. 

\section{Low-energy effective theory near the phase transition}
The results for the histograms of the order parameter $(M_A,M_B)$ naturally 
lead to the formulation of an effective theory with an approximate $SO(2)$ 
symmetry in terms of a unit-vector field 
$\vec{e}(x) = (\cos \varphi(x),\sin\varphi(x))$ representing the direction 
of $(M_A, M_B)$. The action then takes the form 
\begin{gather}
 S[\varphi] \! = \! \int \! d^3x \frac{1}{c} \! \left[ \frac{\rho}{2} \partial_\mu
\varphi \partial_\mu \varphi \! + \! 
\delta \cos^2(2 \varphi) \! + \! \varepsilon \cos^4(2 \varphi)\right],
\end{gather}
where we used $\partial_3 = \partial_{ct}$. Here $\rho$ is the spin stiffness 
and $c$ is the velocity of an emergent pseudo-Goldstone boson. The $\delta$-term
breaks the symmetry down to $\mathbb{Z}(4)$ and leads to a small Goldstone boson 
mass $M c = 2\sqrt{2|\delta|/\rho}$. The last term ensures that the string 
tension remains proportional to $\sqrt{\epsilon \rho}$, even at the phase transition. 
It is thus non-vanishing because in the effective theory the phase transition happens at 
$\delta_c + \epsilon_c = 0$. The fact that $(M_A,M_B)$ is equivalent to $-(M_A,
M_B)$ reduces the emergent symmetry from $SO(2)$ to $\mathbb{R}P(1)$. Therefore only states
invariant against sign changes of $\vec{e}(x)$ belong to the physical Hilbert space.

\begin{figure}[ht]
\begin{center}
 \includegraphics[width=6cm]{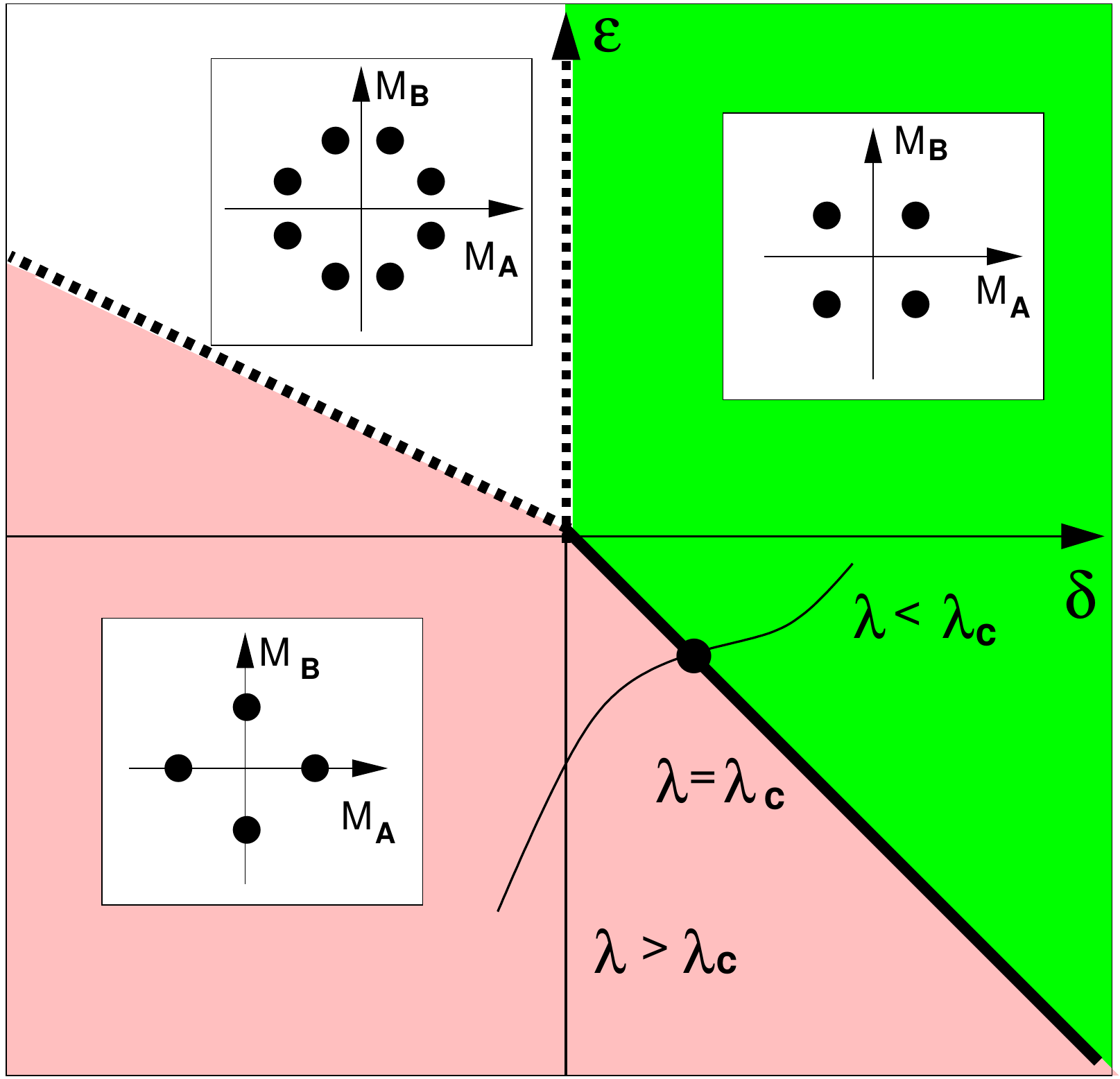}
 \caption{Phase diagram as a function of $\delta$ and $\varepsilon$. The insets indicate the location 
 of the peaks in the distribution $p(M_A,M_B)$. The fat and dashed lines are first and second order 
 phase transitions, respectively. The curved line indicates a possible path taken in the QLM when varying $\lambda$.}
 \label{ET_PhaseDiagram}
\end{center}
\end{figure}

By applying the Ginsburg-Landau-Wilson paradigm to the $\delta$- and $\epsilon$-terms, 
in mean field theory one obtains the phase diagram shown in Fig.\ \ref{ET_PhaseDiagram}. 
The two phases realized in the QLM both have four peaks in the order parameter 
distribution $p(M_A,M_B)$, and are separated by a weak first order phase transition. 
In addition, there is an intermediate phase with eight peaks, separated from the
other phases by second order phase transitions \cite{Ban13a}.

\subsection{Comparison of the effective theory with the exact diagonalization results}
Looking at the energy spectrum obtained by the exact diagonalization calculations
(cf. Fig.\ \ref{statecross}), near $\lambda_c$ we observe an approximate finite-volume rotor 
spectrum $E_m = \frac{m^2c^2}{2\rho L_1L_2}$ for even values of $m$.
In the effective theory at the phase transition we get the same spectrum. Analyzing the eigenstates for 
their quantum numbers, we obtain $C = +, p = (0,0)$ for the ground state $m = 0$, 
$C = \pm, p = (\pi,\pi)$ for the next two states $m = \pm 2$ and $C = \pm, p = (0,0)$
for the $m = \pm 4$ states. All of this is consistent with the spectrum at
$\lambda_c$ shown in Fig.\ \ref{statecross}.

\begin{figure}[ht]
\begin{center}
 \includegraphics[width=13cm]{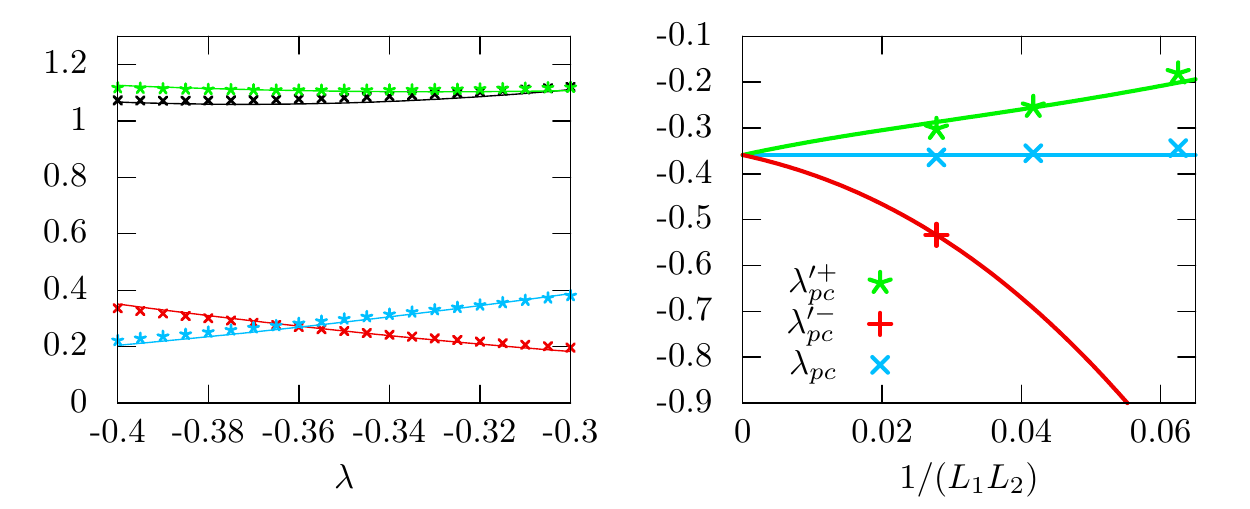}
 \caption{Global fit of the energy gaps near $\lambda_{pc}$ for 
$E_{\pm}$ and $E'_{\pm}$ (left) and of the $L_1L_2$-dependence of
the crossing points $\lambda_{pc}$, $\lambda'^+_{pc}$, $\lambda'^-_{pc}$ (right).}
 \label{GlobalFit}
\end{center}
\end{figure}

By expanding in powers of $\delta L_1L_2$ and $\epsilon L_1 L_2$, we obtain predictions 
for the energy gaps as functions of the parameters,
\begin{align}
 E_+ &= \frac{2 c^2}{\rho L_1L_2} + \frac{L_1L_2}{4}(\delta + \epsilon) + \frac{\rho
L_1^3 L_2^3}{256 c^2}\left( 3\delta^2 + \frac{11}{2}\delta\epsilon +
\frac{119}{48}\epsilon^2\right) + O(\epsilon^3,\, \delta^3) \\
 E_- &= \frac{2 c^2}{\rho L_1L_2} - \frac{L_1L_2}{4}(\delta + \epsilon) + \frac{\rho
L_1^3 L_2^3}{256 c^2}\left( 3\delta^2 + \frac{13}{2}\delta\epsilon +
\frac{167}{48}\epsilon^2\right) + O(\epsilon^3,\, \delta^3) \\
 E'_+ &= \frac{8 c^2}{\rho L_1L_2} + \frac{L_1L_2}{16} \epsilon + \frac{\rho L_1^3
L_2^3}{128 c^2}\left( \frac{11}{3}(\delta+\epsilon)^2 +
\frac{3}{128}\epsilon^2\right) + O(\epsilon^3,\, \delta^3) \\
 E'_- &= \frac{8 c^2}{\rho L_1L_2} - \frac{L_1L_2}{16} \epsilon + \frac{\rho L_1^3
L_2^3}{128 c^2}\left( \frac{5}{3}(\delta+\epsilon)^2 +
\frac{3}{128}\epsilon^2\right) + O(\epsilon^3,\, \delta^3)  \\
 E''_+ &= \frac{18 c^2}{\rho L_1L_2} + \frac{\rho L_1^3 L_2^3}{512 c^2}\left(
9\delta^2 + 19\delta\epsilon + \frac{409}{40}\epsilon^2\right) + O(\epsilon^3,\,
\delta^3)  \\
 E''_- &= \frac{18 c^2}{\rho L_1L_2} + \frac{\rho L_1^3 L_2^3}{512 c^2}\left(
9\delta^2 + 17\delta\epsilon + \frac{329}{40}\epsilon^2\right) + O(\epsilon^3,\,
\delta^3), 
\end{align}
where the notation is the same as in Fig.\ \ref{statecross} and $E''_{\pm}$ refer to
the energy gaps of the next excited states. From the exact diagonalization results 
we can also extract the level crossing points $\lambda_{pc}$ (crossing of $E_+$ 
and $E_-$), $\lambda'^+_{pc}$ and $\lambda'^-_{pc}$ (the two crossings of $E'_+$ and
$E'_-$). According to the effective theory, they should behave as
\begin{gather}
 \lambda_{pc} = \lambda_c + \frac{A}{(L_1L_2)^2} + O(\frac{1}{L_1^3L_2^{3}}), \\
 \lambda'^{\pm}_{pc} = \lambda_c \pm \frac{1}{L_1L_2}\sqrt{-\frac{8}{c_1c_2^2} (c_3
+ \lambda_c c_4) + \frac{16 c_4^2}{c_1^2c_2^4 L_1^2L_2^2}} - \frac{4
c_4}{c_1c_2^2L_1^2L_2^2} + O(\frac{1}{L_1^3L_2^{3}}),
\end{gather}
where we used the representations $\frac{\rho}{c^2} = c_1$, $\delta + \epsilon =
c_2(\lambda - \lambda_c)$ and $\epsilon = (c_3 + \lambda c_4)$. Both the energy gaps
 and the behavior of the different $\lambda_{pc}$ are in quantitative agreement with 
the exact diagonalization results. A global fit yields 
$\lambda_c = -0.359(5)$, $\delta_c = -\epsilon_c = 0.01(1) \, J/a^2$, $\rho =
0.45(3)\, J$ and $c = 1.5(1)\, Ja$, where $a$ is the lattice spacing. Fig.\ 
\ref{GlobalFit} shows two exemplary comparisons of the fitted functions with 
the values extracted from exact diagonalization.

\section{Conclusion}
We have shown that even the simplest quantum link model has highly non-trivial
physics involving multi-stranded confining strings and an emergent $SO(2)$ symmetry. The 
quantum link model studied here is very closely related to a class of models studied
in condensed matter physics with connections to high-$T_c$ superconductivity, 
known as the quantum dimer models. Our methods and numerical algorithms can be
straightforwardly extended to the dimer model. The corresponding investigation 
is in progress. Our results also encourage the application of dualization
techniques to quantum Hamiltonians for other theories, and, in particular, 
to Hamiltonians of quantum link models in higher dimensions. The development of a 
quantum simulator using optical lattices to study the dynamical features 
of this model would be a very welcome and non-trivial step on the road 
to quantum simulate QCD.

\end{document}